\documentclass[10pt]{article}
\usepackage[cp866]{inputenc}
\begin{document}

\centerline {\Large\bf The noncommutativity of the conservation laws: }
\centerline {\Large\bf Mechanism of origination of vorticity and }
\centerline {\Large\bf turbulence}
\centerline {L.I. Petrova}
\centerline{{\it Moscow State University, Russia, e-mail: ptr@cs.msu.su}}

\begin{abstract}

From the equations of conservation laws for energy, linear
momentum, angular momentum and mass the evolutionary relation in
differential forms follows. This relation connects the differential 
of entropy and the skew-symmetric form, whose coefficients depend on the
characteristics of gas-dynamic system and the external actions.

The evolutionary relation turns out to be nonidentical that is explained by 
the noncommutativity of conservation laws. 

The properties of such nonidentical relation (selfvariation, 
degenerate transformation) enable one to disclose the mechanism 
of evolutionary processes in gas-dynamic system that are 
accompanied by origination of vorticity and turbulence. In this case 
the intensity of vorticity and turbulence is defined by 
the commutator on unclosed skew-symmetric form in the nonidentical 
evolutionary relation. 

\end{abstract}
\bigskip

{\large {\bf Introduction}}

Before passing to the analysis of the conservation laws for the gas-dynamic system
under consideration, it should be told a little of the concept of 'conservation 
laws'.

Owing to the development of science the concept of 'conservation laws' 
has assumed a different meaning in various branches of physics and 
mechanics.

In areas of physics related to the field theory and in the theoretical
mechanics 'the conservation laws' are those according to which there
exist conserved physical quantities or objects. These are the
conservation laws that above were named 'exact'. Such conservation laws 
are described be the closed exterior skew-symmetric forms [1]. 
(It is known that the differential of closed exterior form equals zero, 
that is, the closed form is a conserved quantity).

In mechanics and physics of continuous media the concept of 'conservation laws'
relates to the conservation laws for energy, linear momentum, angular momentum,
and mass that establish the balance between the change of physical quantities
and external action. These conservation laws can be named the balance conservation laws.
They are described by differential (or integral) equations. It may be pointed that all
continuous media such as thermodynamic,
gas dynamical, cosmic systems and others (which can be referred to
as material systems), are subject to the balance conservation laws.

The analysis of the equations of balance conservation laws carried
out using the skew-symmetric differential forms showed that the
balance and exact conservation laws are related to each other. The
conserved physical quantities or objects, whose availability
points out to that the exact conservation laws obey, are obtained
from the equations describing the balance conservation laws. The
process of passing from the balance conservation laws to exact
conservation laws is accompanied by the origination of a certain
physical structures and the emergence in material (continuous)
media of observed formations (fluctuations, turbulent pulsations,
waves, massless particles and others) [2]. The processes of
origination of the vorticity and turbulence are examples of such
processes.

As the analysis of the equations of the balance conservation laws showed, these processes
are the results of noncommutativity of the balance conservation laws.

\section{Gas dynamical system of ideal gas}

First of all, we consider the simplest gas dynamical system,
namely, a flow of ideal (inviscous, heat nonconductive) gas.
In the second part of the paper the gas dynamic system of the
viscous heat conductive gas will be considered.

\subsection{The equation of the conservation law. Evolutionary relation }

Assume that the gas is a thermodynamic system in the state of local equilibrium
(whenever the gas dynamic system itself may be in nonequilibrium
state), that is, the following relation is fulfilled [3]:
$$Tds\,=\,de\,+\,pdV \eqno(1)$$
where $T$, $p$ and $V$ are the temperature, the pressure and the gas  volume,
$s$, $e$ are entropy\index{entropy} and internal energy per unit volume.
[Relation (1) determines the entropy $s$ as a thermodynamical state
function. For the gas dynamical system the thermodynamical state function
describes only the state of the gas dynamical element (a gas particle).  For the
gas dynamical system the state function is also the entropy. But in  this case
the entropy is a function of space-time coordinate.]

Let us introduce two frames of reference: an inertial one that is not connected
with the gas dynamical system and an accompanying frame of reference that is connected
with the manifold formed by the trajectories of the  system elements.
(Both Euler's and Lagrange's systems of coordinates
can be examples of such frames [4]).

In the inertial frame of reference the Euler equations are the
conservation laws for energy, linear momentum and mass of ideal gas [4].

The equation of the conservation law of energy for ideal gas can
be written as
$${{Dh}\over {Dt}}- {1\over {\rho }}{{Dp}\over {Dt}}\,=\,0 \eqno(2)$$
where $D/Dt$ is the total derivative with respect to time (if to designate
the spatial coordinates by $x_i$ and the velocity components by $u_i$,
$D/Dt\,=\,\partial /\partial t+u_i\partial /\partial x_i$). Here  $\rho=1/V $
and $h$ are respectively the mass and enthalpy densities of the gas.

Expressing enthalpy in terms of internal energy $e$ with the help of formula
$h\,=\,e\,+\,p/\rho $ and using relation (1), the balance conservation law
equation can be put to the form
$${{Ds}\over {Dt}}\,=\,0 \eqno(3)$$

And respectively, the equation of the conservation law for linear
momentum can be presented as [4,5]
$$\hbox {grad} \,s\,=\,(\hbox {grad} \,h_0\,+\,{\bf U}\times \hbox {rot} {\bf U}\,-{\bf F}\,+\,
\partial {\bf U}/\partial t)/T \eqno(4)$$
where ${\bf U}$ is the velocity of the gas particle,
$h_0=({\bf U \cdot U})/2+h$, ${\bf F}$ is the mass force. The operator $grad$
in this equation is defined only in the plane normal to the trajectory.
[Here it was tolerated a certain incorrectness. Equations (3), (4) are
written in different forms. This is connected with difficulties when
deriving these equations themselves.
However, this incorrectness will not effect on results of the qualitative
analysis of the evolutionary relation obtained from these equations.]

Since the total derivative with respect to time is that along the trajectory,
in the accompanying frame of reference equations (3) and (4)
take the form:
$${{\partial s}\over {\partial \xi ^1}}\,=\,A_{1} \eqno (5)$$
$${{\partial s}\over {\partial \xi ^{\nu}}}\,=\,A_{\nu },\quad \nu=2, ... \eqno(6)$$
where $\xi ^1$ is the coordinate along the trajectory, $A_{1}=0$ (see
equation (3)), $\partial s/\partial \xi ^{\nu }$ is the left-hand side
of equation (4), and $A_{\nu }$ is obtained from the
right-hand side of equation (4).

Equations (5) and (6) can be convoluted into the relation
$$ds\,=\,\omega \eqno(7)$$
where $\omega\,=\,A_{\mu} d\xi ^{\mu}$ is the first degree differential form
(here $\mu =1,\,\nu $).

Since the equations of conservation law are evolutionary ones,
the relation obtained is also an evolutionary relation.

\subsection{Analysis of the evolutionary relation. Nonidentity of the
evolutionary relation.}

The evolutionary relation (7) is a nonidentical one as it involves
the unclosed differential form.

While describing actual processes, the evolutionary form $\omega $ is
not closed. The differential of evolutionary form $\omega $ and
its commutator are nonzero.
The differential of evolutionary
form $\omega $ is expressed
as $d\omega=\sum K_{1\nu}d\xi ^1 d\xi ^{\nu }$, where
$K_{1\nu }$ are the components of the form commutator.
The commutator of differential form $\omega$ is nonzero.
Without accounting for terms that
are connected with the deformation of the manifold formed by
the trajectories, the commutator can be written as
$$K_{1\nu }\,=\,{{\partial A_{\nu }}\over {\partial \xi ^1}}\,-\,{{\partial A_1}\over
{\partial \xi ^{\nu }}}$$

The coefficients $A_{\mu }$ of the form $\omega $ have been obtained
either from the equation of the balance conservation law for energy
or from that for linear momentum. This means that in the first case
the coefficients depend on the energetic action and in the second case
they depend on the force action.
In actual processes energetic and force actions have different nature
and appear to be inconsistent. The commutator of the form $\omega $
constructed of the derivatives of such coefficients is nonzero.
This means that the differential of the form $\omega $
is nonzero as well. Thus, the form $\omega$ proves to be unclosed and
is not a differential. In the left-hand side
of  relation (7) it stands a differential, whereas in the right-hand
side it stands an unclosed form that is not a differential.
Such a relation cannot be an identical one.

The nonidentity of the evolutionary relation means that the
balance conservation law equations are inconsistent. And this indicates
that the balance conservation laws are
noncommutative. (If the conservation laws be commutative,
the equations would be consistent and the evolutionary relation would
be identical).

The evolutionary relation obtained from the balance conservation laws
reflects the character of interactions of the balance conservation laws.
The nonidentity of the evolutionary equation means that the balance
conservation laws are noncommutative, that is, the results of action of
the conservation laws depend on the order in what they act.

{\footnotesize \{It should be noted that for all material systems (such as
thermodynamic, gas dynamical, cosmic  and others) the evolutionary relation obtained
from the equations of balance conservation laws is nonidentical, and this points to
the noncommutativity of balance conservation laws.

The 'noncommutativity' of the balance conservation laws
can be explained in the following manner.

Suppose that firstly the energetic and then the force
perturbations act onto a local domain of the material system (an element
and its neighborhood). Let the local domain be in some state $A$
in the initial instant. According to the balance conservation law for
energy, under exposure to the energetic perturbation the local domain
develops from the state $A$ into the state $B$. Then, according to
the balance conservation law for momentum, under exposure to the force
perturbation it develops from the state $B$ to the state $C$. Suppose
now that the sequence of the actions is changed, namely, firstly the force
perturbation and then the energetic one act, and the system develops
firstly into any state $B'$ and then it proceeds into the state $C'$.
If the state $C'$ coincides with the state $C$ (this corresponds to
the local equilibrium state of the system), that is, the result
does not depend on the sequence of perturbations of different types (and
on the sequence of implementing the relevant balance conservation laws), then
this means that the balance conservation laws are commutative.
If the state $C'$ does not coincide with the state $C$ (that is,
the system state turns out to be not the equilibrium one), this
means that the balance conservation laws prove to be noncommutative.

The reason for noncommutativity of the balance conservation laws is
connected with the fact that the material system is subject to
actions of different nature, the nature of these actions is
inconsistent with the nature of the material system.\}}

Thus, if we even doesn't know the specific expression of the form
$\omega $, one can state that, due to inconsistence of external
actions, the evolutionary relation turns out to be nonidentical
one for real processes, and this points out to the
noncommutativity of balance conservation laws.

To what results the noncommutativity of balance conservation laws leads?

\subsection{Nonequilibrium of the gas dynamical system.}

The further analysis of the  evolutionary relation allows to disclose the
effect of noncommutativity of the balance conservation laws on the evolutionary
processes in gas dynamical system that lead to development of instability and
origination of vorticity.

The role of the evolutionary relation in evolutionary process is
due to the fact that this relation includes the differential of
entropy $ds$, which specifies the state of gas dynamical system.
However, here there is a subtle point. One can obtain the
differential $ds$ from the evolutionary relation only if this
relation proves to be identical. When the relation (7) appears to
be identical one (if the form $\omega $ be a closed form, and
hence it is a differential), one can obtain the differential of
entropy $s$ and find entropy as a function of space-time
coordinates. It is precisely the entropy that will be the gas
dynamic function of state. The availability  of the gas dynamic
function of state would point to the equilibrium state of gas
dynamic system. If relation (7) be not identical, from this
relation the differential of entropy $s$ cannot be defined. This
will point to an absence of the gas dynamic function of state and
nonequilibrium state of the system. {\footnotesize \{It should be
noted once more that in the relation (1), which describes the
thermodynamic system state, the dependence of entropy on
thermodynamical variables is considered, whereas in the
evolutionary relation for gas-dynamic system the dependence of
entropy on space-time coordinates is analyzed. The entropy, which
depends on thermodynamic variables, is a state function of
thermodynamic system, and the entropy, which depends on space-time
coordinates, is a state function of gas dynamic system. In the gas
dynamic system the entropy as a thermodynamic function specifies
the state of gas rather then of the gas dynamic system.\}}

\bigskip
It has been shown above that, since the evolutionary relation is not identical
because of the noncommutativity the conservation laws, from this
relation one cannot get the state differential $ds$  that may point out
to the equilibrium state of the gas dynamical system.
This means that the gas dynamical system state is nonequilibrium.

The nonequilibrium is produced by internal forces that are described
by the commutator of the form $\omega $. (If the evolutionary form
commutator be zero, the evolutionary relation would be identical, and
this would point out to the equilibrium state, i.e. the absence of internal
forces.) Everything that gives the contribution into the evolutionary
form commutator leads to emergence of the internal force that causes
the nonequilibrium state and leads to development of instability.

It becomes evident that a cause of the gas dynamic nonequilibrium and
instability is something that contributes into the commutator of
the form $\omega $.

From the analysis of the expression $A_{\nu }$ and with taking into account
that $A_1\,=\,0$, one can see that the terms related to the multiple
connectedness of the flow domain (the second term of equation (3)),
the nonpotentiality of the external forces (the third term in (3)) and the
nonstationarity of the flow (the forth term
in (3)) contribute to the commutator. All these factors lead to the emergence
of internal forces, the nonequilibrium state and the development of instability.

One can see that the development of instability is caused by the not simple
connectedness of the flow domain,  the nonpotential external
(for each local domain of the gas dynamic system) forces and  the nonstationarity
of the flow. (In the common case, the thermodynamic, chemical, 
oscillatory, rotational and translational nonequilibrium will effect
on the gas dynamic instability).

All these factors lead to emergence of internal forces,
that is, to nonequilibrium and to development of various types of instability.
(It may be noted that, for the case of
ideal gas, Lagrange [4] derived the condition of the eddy-free stable flow.
This condition is as follows: the domain must be simple connected one,
forces must be potential and the flow must be stationary. One can see,
that under fulfillment of these conditions there are no terms that
contribute into the commutator).

And yet, for every type of instability one can find an appropriate term
giving contribution into the evolutionary form commutator, which is
responsible for this type of instability.
Thus, there is the unambiguous connection between the type of instability
and the terms that contribute into the evolutionary form commutator in
the evolutionary relation. \{In the general case one has to consider the
evolutionary relations that correspond to the balance conservation laws
for angular momentum and mass as well\}.

Hence one can see that the noncommutativity of the balance conservation laws
leads to emergence of internal forces (whose value is described by the
evolutionary form commutator) and to appearance of the nonequilibrium.

\bigskip
The nonidentical evolutionary relation is selfvarying one. (Since
one of the objects is an unmeasurable quantity, the other cannot be
compared with the first  one, and hence, the process of mutual
variation cannot stop.)

Such selfvarying of evolutionary relation points to the fact that the gas
dynamic system state changes.
However, in this case the gas dynamic system state remains nonequilibrium
because the nonidentity of evolutionary relation holds.

Whether the gas dynamic system can get rid of
the internal force and transit into the equilibrium state?

\subsection{Transition of the gas dynamic system into a locally equilibrium
state. Origination of physical structures.}

It turns out that the gas dynamic system can transit into the
locally equilibrium state.

From the properties of nonidentical relation
it follows that under selfvariation of nonidentical relation it can
proceed the degenerate transformation when from the unclosed
evolutionary form the skew-symmetric differential form
closed on some
structure (pseudostructure) can be obtained and the identical relation can be
obtained from nonidentical relation. The degrees of freedom of gas dynamic systems
(translational, rotating, oscillating and others) are the conditions of
degenerate transformation.  The conditions of degenerate transformation
specify the pseudostructures: characteristics, singular
points, envelopes of characteristics and so on.

The realization of the conditions of degenerate transformation leads
to the realization of pseudostructure $\pi$ (the closed dual form)
and formatting the closed inexact form $\omega_\pi$, whose closure 
conditions have the form $d_\pi \omega^p=0,  d_\pi{}^*\omega^p=0$.
On the pseudostructure $\pi$, from
evolutionary relation (7) it is obtained the identical relation
$$
ds_\pi=\omega_\pi\eqno(8)
$$
from which the differential $ds_\pi$ can be obtained. This means that there exists
the state function of gas dynamic system, namely, the entropy
whose availability points to the locally-equilibrium state of the gas
dynamic system. (In the papers on gas dynamics it is assumed that from
equation (5) one can obtain the entropy along the
trajectory. However, it occurs that  entropy as a function
of space-time coordinates cannot exist if the conditions of
degenerate transformations are not satisfied. Entropy as a function of 
the state has to be satisfied both to equation (5) and equation (6). 

Realization of pseudostructure $\pi$  and formatting the closed inexact
form $\omega_\pi$ points to emergence of physical structure, i.e. a certain
conserved object. The characteristics, the singular
points, the envelopes of characteristics and so on
(pseudostructures - the closed dual forms) with conserved quantities
(closed inexact form) are examples of such physical structures.

One can see that the identical relation (8) holds the duality. The left-hand side of this
relation includes the differential $ds_\pi$, which specifies  gas dynamic
system and whose availability points to the locally-equilibrium state
of gas dynamic system. And the right-hand side includes a closed inexact
form $\omega_\pi$, which is a characteristics of physical structures.

This shows that the transition of gas dynamic system into the locally
equilibrium state (under realization of degrees of freedom) is
accompanied by the origination of physical structures.

The origination of physical structure reveals as
a new measurable and observable formation that spontaneously arises
in gas dynamic system.

In gas dynamical system of ideal gas formations that correspond to
emerged physical structures are waves, shock waves, vortices.

(In gas dynamical system of viscous gas, formations that correspond
to emerged physical structures are turbulent pulsations).

Since  the created formation is a result of transition of an
unmeasurable quantity described by the evolutionary form commutator
into a measurable physical quantity, it is evident that the intensity of
the formation created is controlled by the quantity that was stored
by the evolutionary form commutator.

One can see that in gas dynamical system, even in the case of
ideal gas, it can originate the physical structures and relevant
formations that lead to emergence of vorticity.

It should be emphasized once more that the origination of various structures
only proceeds under realization of the conditions of degenerate transformations
that are conditioned by the degrees of freedom such as translational, rotating,
oscillating and others.

The conditions of degenerate transformations are realized as vanishing
some functional expressions such as determinants, Jacobians of
transformations, etc. These conditions specify the integral
surfaces (pseudostructures): the characteristics (the determinant
of coefficients at the normal derivatives vanishes), the singular
points (Jacobian is equal to zero), the envelopes of
characteristics of the Euler equations and so on. Under passing
throughout the integral surfaces the gas dynamic functions or
their derivatives suffer shocks (contact shocks). Below we
present the expressions for calculation of such shocks of
derivatives in the direction normal to characteristics (and to trajectories).

The degenerate transformation is realized as a transition from
the accompanying noninertial frame of reference to the locally inertial
frame of reference, that is, the transition one frame of reference to another
nonequivalent frame of reference.
The evolutionary form and nonidentical evolutionary relation are defined in
the noninertial frame of reference (deforming manifold). But the closed
exterior form and the identical relation are obtained with
respect to the locally-inertial frame of reference (pseudostructure).

Let as analyze which types of instability and what gas dynamic
formations can originate under given external action.

1). \emph{ Shock, break of diaphragm and others}. The instability originates
because
of nonstationarity. The last term in equation (4) gives a contribution
into the commutator. In the case of ideal gas whose flow is described by
equations of the hyperbolic type the transition to the locally equilibrium
state is possible on the characteristics and their envelopes. The
corresponding formations are weak and shock waves.

2).\emph{ Flow of ideal (inviscous, heat nonconductive) gas around bodies
Action of nonpotential forces}. The instability develops because of
the multiple connectedness of the flow domain and a nonpotentiality of the
body forces. The contribution into the commutator comes from the second and
third terms of the right-hand side of  equation (4). Since the gas is ideal
one and $\partial s/\partial \xi ^1=A_1=0$, that is, there is no contribution
into the each fluid particle, the instability of convective type develops.
For $U>a$ ($U$ is the velocity of the gas particle, $a$ is the speed of
sound\index{speed of sound})
the set of equations of the balance conservation laws belongs to the
hyperbolic type, and hence the transition to the locally equilibrium state is
possible on the characteristics and on the envelopes of characteristics
as well, and weak and shock waves are formations of the system.
If $U<a$ when the equations are of elliptic type, such a transition is
possible only at singular points. The formations emerged due to the 
convection are of vortex type. At long acting the large-scale structures
can be produced.

\bigskip

Studying the instability on the basis of the analysis of entropy
behavior was carried out in the works by Prigogine and co-authors [6,7].
In that works the entropy was considered as the thermodynamic function
of state (though its behavior along the trajectory was analyzed).
By means of such state function one can trace the development (in gas
fluxes) of the thermodynamic instability only [6]. To investigate the gas
dynamic instability it is necessary to consider entropy as the gas dynamic
state function, i.e. as a function of the space-time coordinates.
Whereas for studying the thermodynamic instability one has to analyze
the commutator constructed by the mixed derivatives of entropy with respect
to the thermodynamic variables, for studying the gas dynamic instability
it is necessary to analyze the commutators
constructed by the mixed derivatives of entropy with respect to the space-time
coordinates.

\section{Gas dynamical system of viscous gas}

In the case of ideal gas the expression $A_1$ in the equation of energy
(see equations (3) and (5)) is equal to zero. In the case of the viscous
heat-conductive gas the expression $A_1$ will depend on the viscosity and
the heat-conductivity.

\subsection{The equation of the conservation law of energy for viscous
gas }

The expression $A_1$ of the equation of energy of the viscous
heat-conductive gas described in the inertial frame of reference
by a set of the Navier-Stokes equations can be written as [4]
$$A_1\,=\,{1\over {\rho }}{{\partial }\over {\partial x_i}}
\left (-{{q_i}\over T}\right )\,-\,{{q_i}\over {\rho T}}\,{{\partial T}\over {\partial x_i}}
\,+{{\tau _{ki}}\over {\rho }}\,{{\partial u_i}\over {\partial x_k}} \eqno(9)$$
Here $q_i$ is the heat flux, $\tau _{ki}$ is the viscous stress tensor.

In the case of viscous gas the terms connected with the transport phenomena
(viscous and heat-conductive) will contribute into the evolutionary
form commutator. This term is responsible for emergence of turbulent
pulsations.

In the general case, the expression $A_1$ will include the terms accounting for
the chemical, oscillatory, rotational, translational and other effects [3 that will
contribute into the evolutionary form commutator and influence on the development of
instability.

Let as analyze the following example.

\emph{ Boundary layer}. The instability originates due to the multiple
connectness of the domain and the transport phenomena (the effect of
viscosity and thermal conductivity). Contributions into the commutator produce
the second term in the right-hand side of equation (4) and the second and
third terms in expression (9). The transition to the locally equilibrium
state is allowed at singular points because in this case
$\partial s/\partial \xi^1=A_1\neq 0$, that is, the external exposure acts
onto the gas particle separately, the development of instability and the
transitions to the locally equilibrium state are allowed only in
the individual fluid particle. Hence, the formations emerged behave as
pulsations. These are turbulent pulsations.

It is commonly believed that the instability is an emergence of
any structures in the gas dynamic flow. From this viewpoint the laminar
boundary layer is regarded as stable one, whereas the turbulent layer
regarded as unstable layer. However the laminar boundary layer cannot
be regarded as a stable one because of the fact that due to the not
simple connectedness of the flow domain and the transport processes
the instability already develops although any formations  do not
originate. In the turbulent boundary layer the emergence of pulsations
is a transition to the locally equilibrium state, and the pulsations
themselves are local formations. The other matter, due to the global
nonequilibrium the locally equilibrium state is broken up and the
pulsations weaken.

\bigskip

Below we present an example of calculating the breaks of derivatives of
the gas dynamic functions that are necessary for numerical analysis of gas
dynamic flows.

\subsection*{Breaks of normal derivatives on characteristics and trajectories}

While studying the effects connected with the origination of the
vorticity one can notice a certain specifics of numerical solving
the Euler equations [4]. This may be demonstrated by the following
example. Assume, that the initial conditions correspond to the
isentropic flow, that is, entropy is the same along all
trajectories. For ideal gas under consideration the entropy
conserves along the trajectory. From this it follows that entropy
has to conserve during all time of flow. However, in actual cases
(unsteady flow, flow along the body, heterogeneous medium) the
derivative of entropy along the direction normal to trajectory
suffers the break. Thus we have that, from one side, entropy
(function) must be constant and, from other hand, its derivatives
suffer the breaks. This contradiction is resolved with taking into
account the fact that the break of derivative is compensated by
changing the stream function or bending the trajectory. It is this
effect that must be accounted for in the process of numerical
calculation.  In particular, when calculating the one-dimensional
nonstationary nonisentropic flow of gas, the conditions on the
characteristics includes the derivative of entropy with respect to
the coordinate normal to trajectory (in space of two variables,
namely, time and coordinate). To calculate this derivative one
must know the break of derivative of entropy. This can be obtained
from the relations that connects the breaks of derivatives of the
gas dynamic functions.

These relations are found from the dynamic conditions of the consistency of
the Euler equations. In paper [8] the dynamical conditions of consistency
of the Euler equations for the case $p=f(\rho)$ were considered.

In the present work by a similar manner it is analyzed the case
when $p=f(\rho ,s)$, where $s$ is the entropy, and the relations
that connect the breaks of derivatives of the functions describing
the particle velocity, the sound speed, and entropy are obtained.
These relations enable one to carry out numerical calculations of
the nonisentropic gas flows.

The scheme of obtaining these relations for one-dimensional nonstationary
equations is the following. At the beginning, the Euler equations are written
down. Then the equations for characteristics and the conditions on
characteristics are derived. The kinematic conditions of consistency [8],
which mutually
connect the breaks of derivatives of the gas dynamic functions, are
written down. These
conditions are substituted into the Euler equations. As a result, the
homogeneous set of equations for the breaks of derivatives of the
functions desired is obtained. On the characteristic surface the
determinant of this set equals zero, and from this it is found the
nontrivial solution for the breaks of derivatives of the functions desired
in their dependence on a value of one of others.

If to take $U$ (the gas velocity), $a$(the sound speed), and $s$,
the following relations are obtained:

1) In the direction normal to the trajectory
the derivatives of the sound speed and entropy suffer breaks (the derivative
of velocity does not suffer a break). These breaks are connected between
them by the relation:
$$\left [{{\partial a}\over {\partial \eta _1}}\right ]=\left [{{\partial s}\over
{\partial \eta _1}}\right ] {a\over {2\gamma s}}$$
where $\eta _1$ is the direction normal to the trajectory, $\gamma$
is the Poisson constant.

2) In the direction normal to the characteristics the derivatives of the
gas velocity and the speed of sound suffer breaks (the derivative of
entropy does not suffer break). These breaks are connected between them
by the relation:
$$\left [{{\partial u}\over {\partial \eta _{+-}}}\right ]= \pm \left [{{\partial a}
\over {\partial \eta _{+-}}}\right ]{2\over{\gamma -1}}$$
where $\eta _{+-}$ are the directions normal to the corresponding
characteristics.

\bigskip

In conclusion it should be said a little about modelling instable flows.
As it is known, some authors tried to account for the development of
instability by means of improving the equations modelling the balance
conservation laws (for example,
by introducing the high-order moments) or by introducing additional
equations. However, such attempts give no satisfactory results. To describe
the nonequilibrium flow and the emergence of the gas dynamic structures
(waves, vortices, turbulent pulsations) one must add
the evolutionary relation obtained from the balance conservation law equations
to the balance conservation law equations.  Under numerical modelling
the gas flows, one has to trace for the transition from the evolutionary
nonidentical relation to the identical relation (the transition from
the evolutionary unclosed form to exterior closed form), and this will
point to emergence of a certain physical structure.

1. Cartan E., Les Systemes Differentials Exterieus ef Leurs Application
Geometriques. -Paris, Hermann, 1945.

2. Petrova L.I., The mechanism of generation of physical structures. 
// Nonlinear Acoustics - Fundamentals and Applications 
(18th International Symposium on Nonlinear Acoustics, Stockholm, Sweden, 
2008) - New York, American Institute of Physics (AIP), 2008, pp.151-154.

3. Haywood R.~W., Equilibrium Thermodynamics. Wiley Inc. 1980.

4. Clark J.~F., Machesney ~M., The Dynamics of Real Gases. Butterworths,
London, 1964.

5. Liepman H.~W., Roshko ~A., Elements of Gas Dynamics. Jonn Wiley,
New York, 1957.

6. Prigogine I., Introduction to Thermodynamics of Irreversible
Processes. --C.Thomas, Springfild, 1955.

7. Glansdorff P., Prigogine I. Thermodynamic Theory of Structure, Stability
and Fluctuations. Wiley, N.Y., 1971.

8. Smirnov V.~I., A course of higher mathematics. -Moscow,
Tech.~Theor.~Lit. 1957, V.~4 (in Russian).

\end{document}